*Low-energy electron transmission imaging of clusters on free-standing graphene*

Jean-Nicolas Longchamp*, Tatiana Latychevskaia*, Conrad Escher, Hans-Werner Fink

Physics Institute, University of Zurich

Winterthurerstrasse 190

8057 Zurich

Switzerland

*Abstract*

We investigated the utility of free-standing graphene as a transparent sample carrier for imaging nanometer-sized objects by means of low-energy electron holography. The sample preparation for obtaining contamination-free graphene as well as the experimental setup and findings are discussed. For incoming electrons with 66 eV kinetic energy graphene exhibits 27% opacity per layer. Hence, electron holograms of nanometer-sized objects adsorbed on free-standing graphene can be recorded and numerically reconstructed to reveal the object's shapes and distribution. Furthermore, a Moiré effect has been observed with free-standing graphene multi-layers.

Corresponding author: longchamp@physik.uzh.ch

* Both authors contributed equally to the paper

*Main text*

The structural analysis of individual molecules, in particular of proteins is the ultimate goal driving the development of novel microscopy techniques. Current pursuits involve X-rays or electrons with wavelengths of the order of one Ångstrom or less. When imaging an individual biological specimen with X-rays or high-energy electrons, the resolution is mainly limited by radiation damage. On the contrary, with low-energy electrons (50-250eV) the infliction of radiation damage is minimal and yet the wavelength sufficiently short (0.7-1.7 Å) [1].

The highest possible resolution is achieved when no lenses, and thus no aberrations, are present in the optical system. One way to circumvent the employment of lenses was suggested by Gabor in 1948 with his invention of holography [2]. For a holographic record the wave scattered off the object under study is superimposed with a well defined reference wave. In the transmission mode, the reference wave is provided by the part of the wave that is not scattered as it passes the object. Thus, to ensure the presence of the reference wave, objects under study ideally must be levitating, be attached to some cantilever (e.g. a carbon nanotube) or, more practical, just rest on a transparent substrate. Holography with low-energy electrons has already been employed for imaging individual molecules such as DNA [3-4], phthalocyaninato polysiloxane [5], tobacco mosaic viruses [6], filamentous bacteriophage [7] and ferritin [8].

Until now, the standard sample preparation procedure for in-line holography was stretching elongated objects over holes in thin films [3-8]. With this geometry, the electron wave passing through the holes provides a well-defined reference wave. The wave that scatters off the object constitutes the object wave. Alternatively, objects might as well be deposited onto a transparent (non-absorbing) substrate. With the discovery of graphene [9] and subsequently developed technologies to isolate individual flakes, an ultimately thin carbon film as sample carrier is now available onto which nanometer-sized objects can be deposited [10]. While graphene is just 0.34 nm thick [11] it appears to be the strongest monolayer ever investigated [12]. With the high-energy Transmission Electron Microscope (TEM) graphene has already been successfully used as substrate for imaging objects such as hydrogen atoms [13], gold nanoparticles [14], $CoCl_2$ nanocrystals [15] and biological molecules, such as TMV [10] and stained DNA [16]. When graphene is subject to low-energy electrons, it still holds transparency and does not experience damage during continuous exposure. In addition, since graphene is electrically conductive, it constitutes a uniform equipotential plane [17] eliminating possible field distortions caused by elongated objects present in previous preparation methods [6]. Holograms of objects deposited on graphene can thus be readily reconstructed by employing classical optical wave propagation theory. With these characteristics graphene excels as a unique sample carrier for low-energy electron holography.

*Sample preparation*

Graphene is by definition a one-atom-thick planar sheet of $sp^2$-bonded carbon atoms. However, its preparation procedures (Chemical Vapor Deposition (CVD), mechanical or chemical exfoliation, epitaxial growth, etc.) always result in some residual contamination on the graphene surface. The chemical nature of these contaminations is directly associated with the materials involved in the process of preparation. The size of the contaminating particles can be of the order of a few nanometers. This usually does not cause any trouble when using graphene for purposes other than imaging at the sub-nanometer scale. However, when using graphene as a sample carrier in electron microscopy, special care must be taken to ensure preparation of clean graphene.

The graphene layers used for our investigation were grown by the CVD method on polycrystalline copper foils [18] and were transferred onto metal coated silicon nitride membranes using standard methods [19]. The membranes were previously perforated with the help of a focused gallium ion beam. The transfer procedure results in clean graphene layers covering holes of 250 and 500 nm in diameter milled through metal coated silicon nitride membranes. Because of the polycrystalline character of the copper substrate, several orientations and multi-layer graphene were anticipated and actually also observed in our experiments and described below.

*Experimental setup*

The experimental setup for holography with low-energy electrons [8] is schematically shown in Fig. 1. An ultra-sharp tungsten tip acts as a coherent point source for low-energy electrons field emitted into vacuum towards the sample. The sample is typically placed at a distance between 200 to 1000 nm from the tip and the detector unit (consisting of a microchannel plate (MCP), a fiber optic plate (FOP) followed by a 16 bit CCD camera) is placed at a distance of 68 mm from the tip. The images are displayed on the 75 mm diameter phosphor coated FOP surface and sampled with 6000x8000 pixel$^2$. The numerical aperture of the setup amounts to N.A.= 0.48.

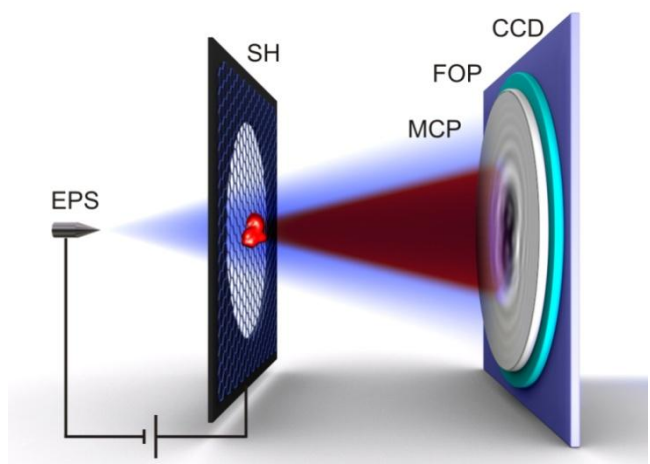

Fig. 1. Schematics of the experimental setup. EPS – electron point source, a very sharp tungsten tip. SH – sample holder, graphene covers a hole of about 500 nm in diameter in a metal-coated silicon nitride membrane. The object under study is located on the graphene substrate. A microchannel plate (MCP), fiber optic plate (FOP) and charge-coupled device (CCD) form the detector unit.

*Transmission through graphene*

It has previously been reported that graphene is extremely transparent to visible light, showing an opacity of 2.3% per single layer [20]. Analogically we find a one order of magnitude higher opacity for low-energy electrons. Fig. 2 shows a hologram of a hole in a metal-coated silicon nitride membrane entirely covered by graphene layers. The hologram is recorded at 66 eV. In the most transparent part of the hologram (Fig. 2(a) top left corner), the only indication of the presence of graphene are small particles resting on its surface. The transmission of the sample within the hole decreases stepwise with an increasing number of graphene layers as shown in Fig. 2(b). We presume (that there is a single layer of graphene in the top left corner and) that each additional layer of graphene absorbs or backscatters the same percentage of the incoming electrons which provides us with the 100% transmission value for the vacuum. The calculations from these data show that graphene imaged with 66 eV energy electrons absorbs or backscatters 27% of the incoming electrons per layer.

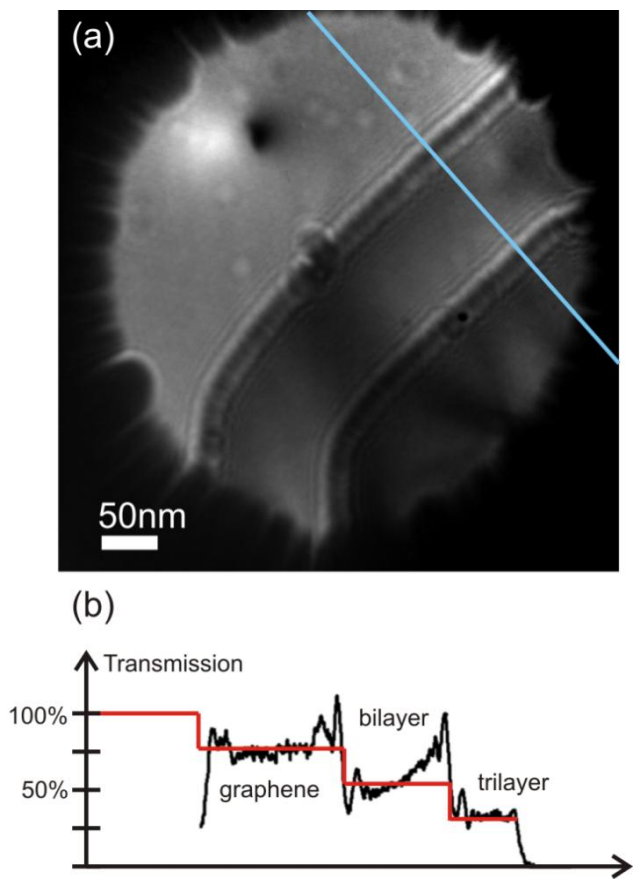

Fig. 2. (a) Hologram of a graphene sample recorded with low-energy electrons at an energy of 66 eV. (b) The intensity profile along the blue line in (a) shows a discrete change in transmission.

*Holograms and their reconstructions*

Holograms of graphene samples recorded with low-energy electrons are shown in Fig. 3. The sample, a metal-coated silicon nitride membrane with holes of 500 nm in diameter covered with graphene, was placed 1070 nm (hologram in Fig. 3(a)) respectively 960 nm (hologram in Fig. 3(b)) in front of the electron source; the kinetic energies of the electrons were 64 and 67 eV respectively. Prior to the numerical reconstruction, the holograms were normalized by division with the background image. The latter was obtained by a two-dimensional Gaussian fit of the selected intensity regions where no interference pattern was observed. The holograms were reconstructed using the back-propagation Fresnel-Kirchhoff integral [21] and the reconstructed amplitude distributions are shown in Fig. 3. Two types of objects can be observed on the graphene surface: extended objects and small particle-like objects. The size of the small particles corresponds to approximately 10 nm.

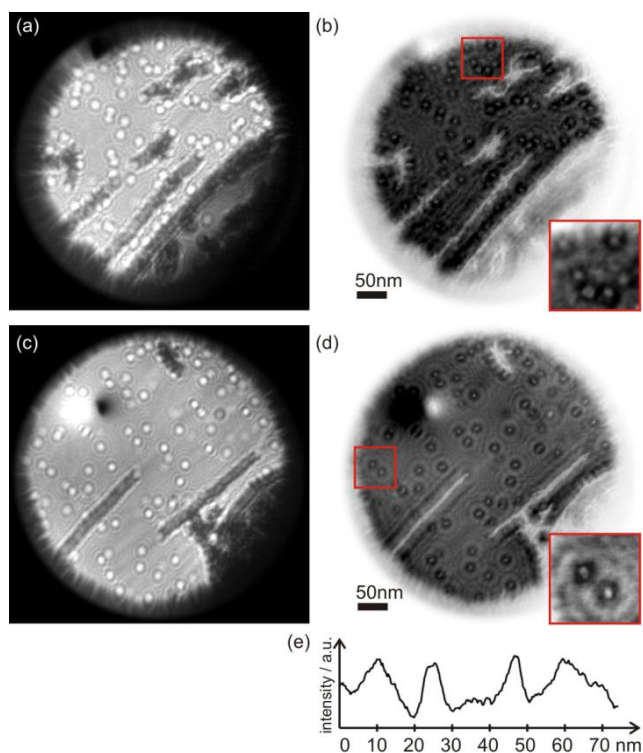

Fig. 3: (a) Hologram recorded with 64 eV electrons and (b) its reconstruction. (c) Hologram recorded with 67 eV electrons and (d) its reconstruction. The insets show the magnified areas in the reconstruction marked by red squares, where the small particles are observed. (e) Intensity profile of the red square in (d) from the top left corner to the right bottom corner, showing two central peaks of about 10 nm in width representing the two objects.

*Moiré*

A Moiré pattern created by the superposition of a few graphene layers was previously observed in STM images of graphene on a substrate [22-26] and in TEM [27]. Here, we report observations of a Moiré pattern in free-standing graphene layers. Figure 4 shows an example of a low-energy electron hologram of graphene recorded with an 800 nm electron source to sample distance; the width of the hologram corresponds to 500 nm. In the lower part of the hologram, shown in Fig. 4(a), a hexagonal periodic structure is apparent. This is the result of the Talbot effect for two superimposed layers of graphene creating a Moiré pattern. Calculations show that the exhibited periodic structure can be attributed to two graphene layers rotated by 2.9° relative to each other, as shown in Fig. 4(b). Another example of a Moiré pattern in free-standing graphene multi-layers is shown in Fig. 5. Here, some additional objects are observed on the graphene surface and reconstructed at a distance of 440 nm from the electron source, as shown in Fig. 5(c). The example in Fig. 5 demonstrates that even two superimposed graphene layers are sufficiently transparent for low-energy electron holography, and that objects deposited on two graphene layers exhibit enough holographic contrast to be reconstructed.

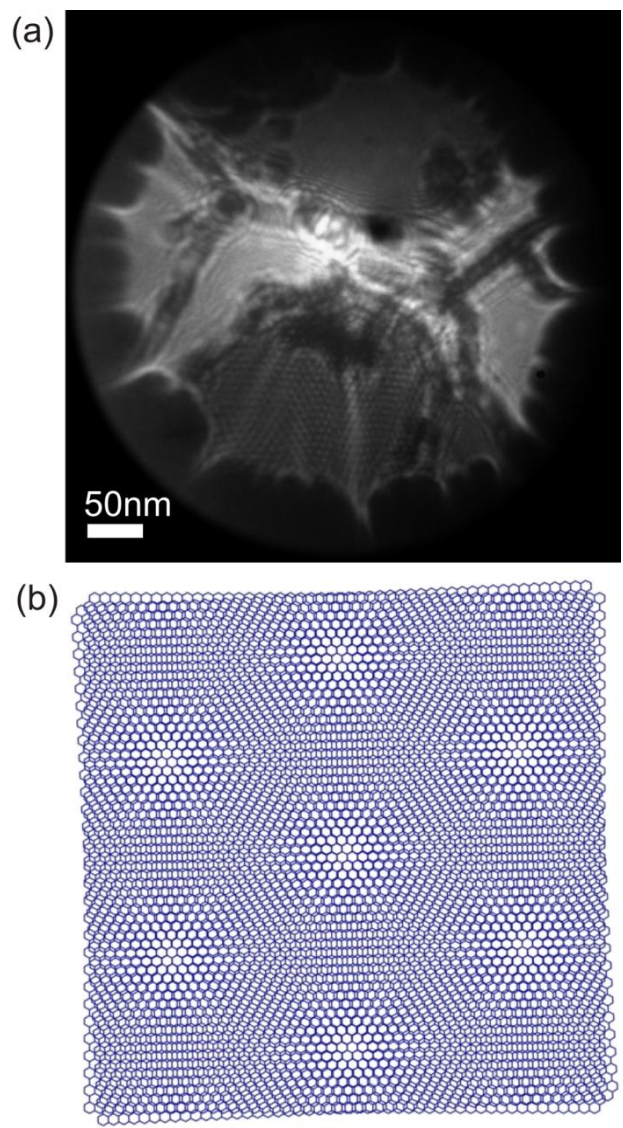

Fig. 4. (a) Hologram of graphene recorded with 58 eV kinetic energy electrons. (b) Drawing of the superposition of two graphene layers rotated by 2.9° relative to each other creating a Moiré pattern matching (a).

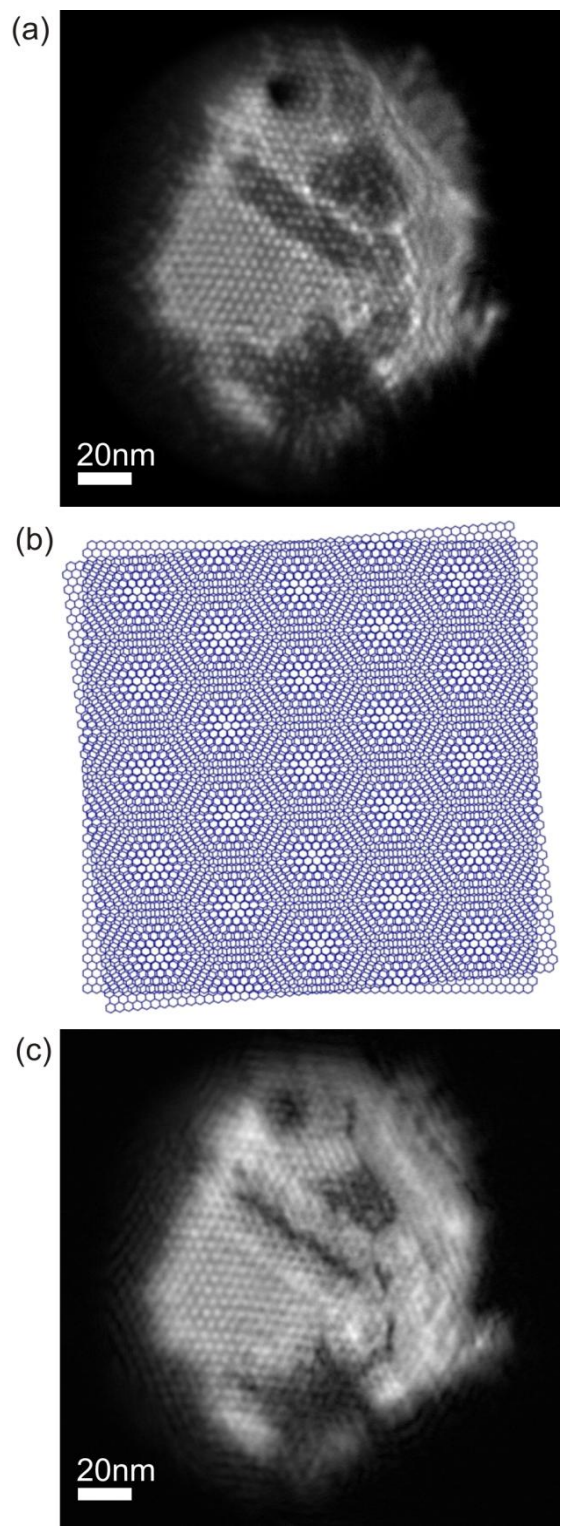

Fig. 5. (a) Hologram of graphene recorded with 55 eV kinetic energy electrons. (b) Drawing of the superposition of two graphene layers rotated by 5.5° relative to each other creating a Moiré pattern matching (a). (c) Reconstruction of the hologram obtained at a distance of 440 nm from the electron source.

*Conclusion*

In summary, we have demonstrated that graphene can be used as sample carrier for nanometer-sized objects to be observed by means of low-energy electron holography. It is reasonably transparent to low-energy electrons with 27% opacity per layer. Furthermore, graphene ensures a well-defined field distribution along the imaging system. As a result, the recorded holograms do not exhibit unwanted artifacts and can easily be reconstructed using classical optical wave propagation integrals. It is expected that the experimental methods presented here can be extended for imaging biomolecules deposited onto free-standing graphene.

*Acknowledgements*

The authors are grateful for financial support by the Swiss National Science Foundation.